\documentclass[prb, longbibliography, twocolumn]{revtex4-2}
\usepackage{bm}
\usepackage{graphicx}
\usepackage{amsmath}
\usepackage{amssymb} 
\usepackage[utf8]{inputenc}
\usepackage[T1]{fontenc}
\usepackage{color}
\usepackage{xcolor}
\usepackage{upgreek} 
\usepackage{subfigure}
\usepackage[unicode=true,colorlinks=true,citecolor=blue]{hyperref}
\usepackage{lipsum}
\usepackage{epsfig}
\usepackage{wrapfig}
\usepackage[normalem]{ulem}
\usepackage{units}
\usepackage{cancel}
\usepackage{float}
\usepackage{lipsum}
\usepackage{placeins}

\newcommand{\nix}[1]{}

\renewcommand{\phi}{\varphi}

\renewcommand{\i}{\mathrm i}
\newcommand{\e}{\mathrm e}
\newcommand{\eps}{\varepsilon}

\newcommand{\beq}{\begin{equation}}
	\newcommand{\eeq}{\end{equation}}
\newcommand\beqa{\begin{eqnarray}}
	\newcommand\eeqa{\end{eqnarray}}
\newcommand\ba{\begin{array}}
	\newcommand\ea{\end{array}}

\begin{document}
	
	\title{Electric and spin-valley currents induced by structured light in 2D Dirac materials}

	\author{A. A. Gunyaga}
	\affiliation{Ioffe Institute, 194021 St. Petersburg, Russia}
	
	\author{M. V. Durnev} 
	\email{durnev@mail.ioffe.ru} 
	\affiliation{Ioffe Institute, 194021 St. Petersburg, Russia}

	\author{S. A. Tarasenko}
	\affiliation{Ioffe Institute, 194021 St. Petersburg, Russia} 
	
\begin{abstract}
Structured optical fields can be used for the injection and control of charge and spin-valley currents. Here, we present a systematical study of these phenomena for interband absorption of structured light in 2D Dirac materials. We derive general expressions for the current density and the quasi-classical generation rate of photoelectrons in the momentum, coordinate, and spin-valley spaces. We reveal mechanisms of the current formation determined by the local and non-local 
contributions to the optical generation, including the mechanisms related to optical alignment of electron momenta by linearly polarized light, optical orientation by circularly polarized light, and the class of charge and spin-valley photon drags sensitive to the phase and polarization profiles of the optical field. We develop a kinetic theory of electric and spin-valley currents driven by the optical field with spatially inhomogeneous intensity, polarization, and phase and obtain analytical expressions for the current contributions. The theory is applied to analyze the photocurrents emerging in TMDC layers and graphene excited by polarization gratings.
\end{abstract}
	
	\maketitle
	 

\section{Introduction}

Interaction of light with atomically thin materials is a fascinating topic of 
research in solid state physics~\cite{Abajo:2025}.
Of quite a broad interest are photoelectric phenomena where light drives charge, spin, or valley currents in 2D Dirac materials, 
such as graphene or transition metal dichalcogenide (TMDC) layers~\cite{Karch2010,Entin2010,Obraztsov2014,Koppens2014,Ivchenko:2008,Karch:2011b,Golub:2011,Muniz:2015,Quereda:2018,Kiemle:2020,Entin:2021}. 
To increase and control the currents one can use 
focused optical beams 
with designed spatial structure of electromagnetic field, the so called structured light~\cite{Forbes2021, Sederberg:2022}. Sophisticated experiments on semiconductor structures and Weyl semimetals 
show the photoresponse sensitive to the orbital angular momentum~\cite{Ji2020,Lai2022,Feng:2022,Session:2025} and the vector structure of light beam~\cite{Sederberg2020}, as well as the nonlocal photoresponse sensitive to the photon helicity~\cite{Karch2011,Wang:2024}. 

The achievements in tailoring the spatial structure of optical field stimulate theoretical studies of photoelectric phenomena in spatially inhomogeneous field.
For intraband excitation of electron gas by structured radiation, the kinetic theory of emerging photocurrents is developed in Refs.~\cite{Gunyaga:2023,Gunyaga:2025},
see also Ref.~\cite{Zuev:2025} for superconducting systems.
It reveals the emergence of textured dc currents and currents at the double frequency, which are fully controlled by the spatial structure of radiation. 
For the spectral range corresponding to the optical transitions between the valence and conduction bands, where the electron-photon
interaction is particularly strong, theoretical results are fragmentary, see recent review~\cite{Quinteiro-Rosen:2022}. They primarily concern
quantum-mechanical calculations of nonstationary currents in quantum
rings and dots~\cite{Quinteiro2009b,Waetzel2016,Waetzel2020} and the rotational photon drag by optical vortices~\cite{Quinteiro2010}.

\begin{figure}[t]
	\centering
	\includegraphics[width=0.95\linewidth]{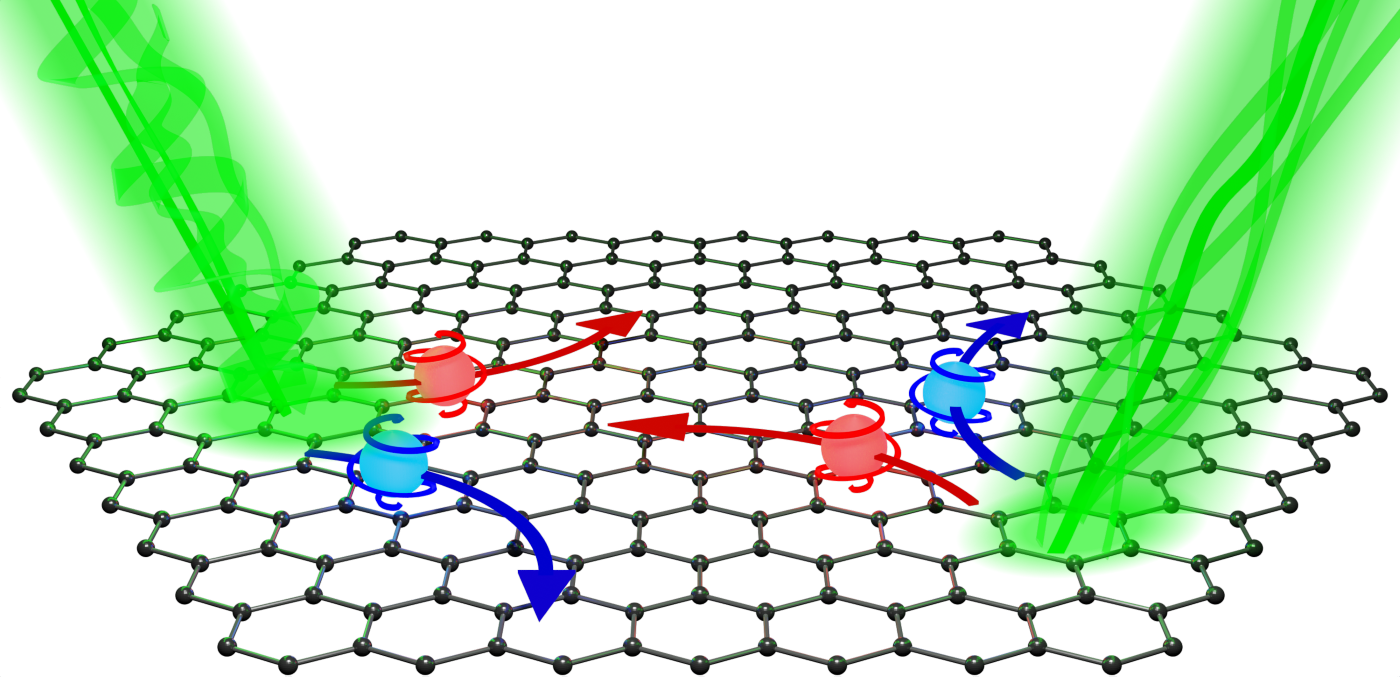}
	\caption{Structured light drives charge and spin-valley currents in 2D Dirac material. The currents are controlled by the spatial structure of electromagnetic field  including the profiles of intensity, polarization, and phase.}
	\label{main_picture}
\end{figure}

Here, we fill this gap and develop a systematic kinetic theory of electric and spin-valley currents induced by structured light at interband optical transitions.
We uncover different microscopic mechanisms of the current generation in 2D Dirac materials with account for relaxation and recombination of photoinduced carriers. As a result, we derive analytical expressions for the contributions to the electric and spin-valley currents driven by the spatial gradients of the optical field intensity, polarization, and phase, see Fig.~\ref{main_picture}.

The paper is organized as follows. In Sec.~\ref{Sect_II} we consider the interband absorption of structured light. Using the $\bm k$$\cdot$$\bm p$ theory and 
the density matrix approach, we derive general expressions for the current density and the quasi-classical generation rate of photoelectrons in the momentum and coordinate spaces.  
The generation term can be presented as the sum of two contributions: the contribution determined by the local amplitude and polarization of the optical field and the non-local contribution determined by spatial gradients of the field and sensitive to the field phase. In Sec.~\ref{Sec_2D}, we describe the band structure 
of 2D Dirac materials and spin-valley selective optical transitions. Sec.~\ref{Sec_currents} presents a kinetic theory of electric and spin-valley currents induced by structured light. We systematically describe the currents related to the Dember effect, optical alignment of electron momenta by linearly polarized light, optical orientation by circularly polarized light, and the class of electric and spin-valley photon drag currents sensitive to the phase profile of the optical field. Additionally, 
the emerging electric and spin-valley currents are converted into each other due to the direct and inverse spin/valley Hall effects.  In Sec.~\ref{Sec_symmetry} the calculated currents are classified based on the symmetry analysis.
The obtained results are discussed in Sec.~\ref{Sec_Results} on the example of polarization gratings formed by linearly or circularly polarized coherent optical fields in TMDC layers and graphene.

\section{General formalism}\label{Sect_II}

\subsection{Spatially varying photocurrents}
	
	We start our paper by deriving general equations for the density of current $\bm j (\bm r)$ induced by spatially inhomogeneous radiation.
We assume that the electron system is described by the matrix Hamiltonian ${\cal H}$ which is linear in the wave vector $\bm k$.
Such a Hamiltonian naturally follows from a multi-band $\bm k$$\cdot$$\bm p$ theory and has the general form ${\cal H} = \beta + \bm \alpha \cdot \bm k$,
where $\beta$ and $\bm \alpha = d {\cal H} /d \bm k$ are the Hermitian matrices, whose dimension coincides  with the number of bands.

In terms of the quantum field operators, the current density operator is given by
\begin{equation}\label{j_gen1}
\bm j (\bm r) = \hat{\psi}^\dag(\bm r) \bm v \hat{\psi} (\bm r) \,,
\end{equation}
where $\hat{\psi} (\bm r) = \sum_{n \bm k} \hat{a}_{n \bm k} \psi_{n \bm k} (\bm r)$, $\hat{a}_{n \bm k}$ are the annihilation operators,
$\psi_{n \bm k} (\bm r) = \chi_{n \bm k} \exp(\i \bm k \cdot \bm r)$, $\chi_{n \bm k}$ are the eigen columns of ${\cal H}$, $n$ is the band index,
and $\bm v = \bm \alpha /\hbar$. Averaging Eq.~\eqref{j_gen1} over the ensemble of electrons and 
introducing the density matrix $\rho_{n\bm k, n'\bm k'} = \langle \hat{a}_{n \bm k}  \hat{a}^\dag_{n' \bm k'}  \rangle$, one obtains the current density 
\begin{equation}\label{j_gen1a}
\bm j (\bm r) = \sum_{n\bm k, n'\bm k'} \rho_{n\bm k, n'\bm k'} \bm v_{n'\bm k', n\bm k} \exp [\i (\bm k - \bm k') \cdot \bm r]  \,,
\end{equation} 
where $\bm v_{n' \bm k', n\bm k} = \chi_{n' \bm k'}^\dag \bm v  \chi_{n \bm k}$.

We focus on the current in the conduction band 
\begin{equation}\label{j_gen2}
\bm j (\bm r) = \sum_{\bm k, \bm k'} \rho^c_{\bm k, \bm k'} \bm v_{\bm k', \bm k}^{c} \exp [\i (\bm k - \bm k') \cdot \bm r]  \,,
\end{equation} 
which is determined by the conduction-band density matrix $\rho^c_{\bm k, \bm k'} = \rho_{c\bm k, c\bm k'}$ and the intraband matrix elements of the velocity operator $\bm v_{\bm k', \bm k}^{c} = \chi_{c \bm k'}^\dag \bm v  \chi_{c \bm k}$. Similar equations can be readily written for the current carried by holes in the valence band. Note that, 
for spatially inhomogeneous radiation, one can also expect an interband contribution to the current~\eqref{j_gen1a} determined by the interband components of the density matrix with $n' \neq n$. This contribution is somewhat similar to the shift contribution to the photogalvanic current in
noncentrosymmetric crystals and to the photon drag current in centrosymmetric crystals~\cite{Belinicher:1982,Shi:2021,Qu:2025}. 
It is small compared to the current studied below and will be considered elsewhere. 

The spatial scale of the current inhomogeneity is defined by the spatial structure of the light beam and is typically much larger than 
the electron wavelength. Then, Eq.~\eqref{j_gen2} can be fruitfully rewritten in the form
\begin{equation}\label{j_gen3}
\bm j (\bm r) = \sum_{\bm k, \bm p} \rho^c_{\bm k + \bm p/2, \bm k - \bm p/2} \bm v_{\bm k - \bm p/2, \bm k + \bm p/2}^{c} \exp (\i \bm p \cdot \bm r)  
\end{equation}
with $|\bm p | \ll |\bm k|$.
Expanding the velocity matrix elements $\bm v_{\bm k - \bm p/2, \bm k + \bm p/2}^{c}$ in the series of $\bm p$, we obtain
\begin{multline}\label{j_gen4}
\bm j (\bm r) = \sum_{\bm k} \bm v_{\bm k}^{c} f(\bm k, \bm r)  \\
+ \frac{\i}{2}  \sum_{\bm k} \sum_{\alpha = x,y} \nabla_{\alpha} f(\bm k, \bm r) \,
 [(\partial_{k_\alpha} - \partial_{k'_\alpha}) \bm v_{\bm k, \bm k'}^{c} ]_{\bm k' = \bm k}  \,,
\end{multline}
where $\bm v_{\bm k}^{c} = \bm v_{\bm k, \bm k}^{c} = (1/\hbar) d \varepsilon_{\bm k}^c / d \bm k$, $\varepsilon_{\bm k}^c$ is
the conduction-band dispersion given by the Hamiltonian ${\cal H}$, 
$f(\bm k, \bm r)$ is the quasi-classical distribution function in the momentum and real spaces
given by the Wigner transformation
\begin{equation}
	f(\bm k, \bm r) = \sum_{\bm p} \rho^c_{\bm k + \bm p/2, \bm k - \bm p/2} \exp(\i \bm p \cdot \bm r) \,,
\end{equation}
and $\nabla = \partial_{\bm r}$ is the spatial gradient.

The conduction-band density matrix $\rho^c_{\bm k, \bm k'}$ and, hence, the quasi-classical distribution function $f(\bm k, \bm r)$
can be calculated by solving the quantum kinetic equation 
\begin{equation}\label{rho_kin}
	\frac{\partial \rho^c_{\bm k, \bm k'}}{\partial t} + \frac{\i}{\hbar} (\varepsilon_{\bm k}^c - \varepsilon_{\bm k'}^c) \rho^c_{\bm k, \bm k'} 
	= g_{\bm k, \bm k'} + I_{\bm k, \bm k'} \{ \rho^c \}\:,
\end{equation}
where $g_{\bm k, \bm k'} $ is the optical generation term and the term $I_{\bm k, \bm k'} \{ \rho^c \}$ describes the processes of
momentum, spin, valley, and energy relaxation and electron-hole recombination.
The Wigner transformation of Eq.~\eqref{rho_kin} gives the kinetic equation for the distribution function $f(\bm k, \bm r)$
\begin{equation}\label{f_kin}
	\frac{\partial f(\bm k, \bm r)}{\partial t} + {\bm v}_{\bm k}^c \cdot  \nabla f(\bm k, \bm r) 
	= g(\bm k, \bm r) + I \{ f \} \,,
\end{equation}
where $g(\bm k, \bm r) =  \sum_{\bm p} g_{\bm k + \bm p/2, \bm k - \bm p/2} \exp(\i \bm p \cdot \bm r)$ is the quasi-classical generation term. 

\subsection{Optical transitions induced by structured radiation}

To proceed further, we calculate the generation term for the interband optical transitions induced by spatially inhomogeneous radiation. 
We take the vector potential of the incident monochromatic field in the $(xy)$ plane of 2D material in the form
\begin{equation}\label{vector_potential}
	\bm A (\bm r, t) = \sum_{\bm q} \bm A_{\bm q} \exp(\i \bm q \cdot \bm r - \i \omega t) + {\rm c.c.} \,, 
\end{equation}
where $\bm A_{\bm q}$ is {the} Fourier amplitude of the vector potential, $\omega$ is the frequency, and $\bm q$ is the in-plane wave vector.
The  generation term in the basis of plane waves has the form~\cite{Ganichev:2003}:
\begin{multline}
	\label{g0}
	g_{\bm k, \bm k'} = \frac{\pi}{\hbar} \sum_{\bm k''} M_{c \bm k, v \bm k''} M_{c \bm k', v \bm k''}^* 
	\\ \times[ \delta(\varepsilon_{\bm k}^c - \varepsilon_{\bm k''}^v - \hbar \omega) + \delta(\varepsilon_{\bm k'}^c - \varepsilon_{\bm k''}^v - \hbar \omega)  ] \,,
\end{multline}
where $M_{c \bm k, v \bm k''}$ is the matrix element of the optical transitions
\begin{equation}\label{matrix_element}
	M_{c \bm k, v \bm k''} = - \frac{e}{c} \sum_{\bm q} \bm v_{cv}(\bm k, \bm k'') \cdot  \bm A_{\bm q} \delta_{\bm k, \bm k'' + \bm q} \,,
\end{equation}
$\bm v_{cv}(\bm k, \bm k'')$ is the interband matrix element of the velocity operator, $\varepsilon_{\bm k}^{v}$ is the valence band energy, 
$\delta(\ldots)$ and $\delta_{\bm k,\bm k''+\bm q}$ are the Dirac delta-function and {the} Kronecker delta, respectively. {At $\bm k = \bm k'$, Eq.~\eqref{g0} yields the well-known Fermi golden rule for generation of the diagonal components of the density matrix $\rho_{\bm k, \bm k'}$.}

The Wigner transformation of Eq.~\eqref{g0} yields
\begin{multline}
	g(\bm k, \bm r) = \frac{\pi e^2}{\hbar c^2} \sum_{\bm q_1,  \bm q_2}\exp(\i \bm p \cdot \bm r) [\bm v_{cv}(\bm k + \bm p/2, \bm k - \bm q) \cdot  \bm A_{\bm q_1} ] \hspace{7em}\\ \hspace{-2em}\times
	[\bm v_{cv}(\bm k - \bm p/2, \bm k - \bm q) \cdot  \bm A_{\bm q_2} ]^* \\[1ex]
	\times \bigl[ \delta(\varepsilon_{\bm k + \bm p/2}^c - \varepsilon_{\bm k - \bm q}^v - \hbar \omega) 
	+ \delta(\varepsilon_{\bm k - \bm p/2}^c - \varepsilon_{\bm k - \bm q}^v - \hbar \omega)  \bigr]  \,,
	\label{g1}
\end{multline}
where $\bm q = (\bm q_1 + \bm q_2)/2$ and $\bm p = \bm q_1 - \bm q_2$. 
Naturally, for the incident radiation in the form of a single plane wave with the in-plane wave vector $\bm q$, Eq.~\eqref{g1} reproduces the known result
$g(\bm k) \propto |\bm v_{cv}(\bm k, \bm k - \bm q) \cdot  \bm A_{\bm q} |^2 \delta(\varepsilon_{\bm k }^c - \varepsilon_{\bm k - \bm q}^v - \hbar \omega)$, see, e.g., Ref.~\cite{Entin2010}.

Using the smallness of the photon wave vectors $\bm q_1$ and $\bm q_2$ compared to the electron wave vector $\bm k$, 
we can expand $g(\bm k, \bm r)$ in the series of $q/k$ as follows
\begin{equation}\label{Eq:g}
	g(\bm k, \bm r) = g^{(0)}(\bm k, \bm r) + g^{(1)}(\bm k, \bm r) + \ldots  
\end{equation}
The term $g^{(0)}(\bm k,\bm r)$ is obtained by neglecting the wave vectors $\bm q_1$ and $\bm q_2$ in the matrix elements 
of the interband velocity operator and in the band energies in Eq.~\eqref{g1}. This gives 
\begin{align}\label{glocal}
	g^{(0)}(\bm k, \bm r) =  \frac{ 2 \pi e^2}{\hbar c^2}  |\bm v_{cv}(\bm k) \cdot  \bm A(\bm r) |^2
	\delta(\varepsilon_{\bm k}^c - \varepsilon_{\bm k}^v - \hbar \omega)  \,,
\end{align} 
where $\bm A(\bm r) = \sum_{\bm q} \bm A_{\bm q} \exp(\i \bm q \cdot \bm r)$ and $\bm v_{cv}(\bm k) = \bm v_{cv}(\bm k, \bm k)$. 
Equation~\eqref{glocal} presents the generation term in the local approximation, where the generation rate at the point $\bm r$ 
is determined by the field at the same point.

The first-order correction $g^{(1)}(\bm k, \bm r)$ is given by
\begin{widetext}
	\begin{align}
		g^{(1)}(\bm k, \bm r) = \frac{2 \pi e^2}{\hbar c^2} \delta({\varepsilon_{\bm k}^g} - \hbar \omega) \, {\rm Im}
		\left \{ [\bm v_{cv}(\bm k) \cdot  \bm A(\bm r)]^* [(\partial_{\bm k}- \partial_{\bm k'}) \cdot \nabla]  
		[\bm v_{cv}(\bm k, \bm k') \cdot  \bm A(\bm r)] \right \}   \nonumber \\
		- \frac{2 \pi e^2}{\hbar c^2}  \delta(\varepsilon_{\bm k}^g  - \hbar \omega) \, {\rm Im}
		\left \{{\partial_{\bm k} \, [\bm v_{cv}(\bm k)} \cdot  \bm A(\bm r)]^* \cdot  
		\nabla \, [\bm v_{cv}(\bm k) \cdot  \bm A(\bm r)] \right \} \nonumber \\
		+ \frac{2 \pi e^2}{c^2} \delta'(\varepsilon_{\bm k}^g  - \hbar \omega) \, {\rm Im}
		\left \{ [\bm v_{cv}(\bm k) \cdot  \bm A(\bm r)]^*  (\bm v_{\bm k}^v \cdot \nabla) \, [\bm v_{cv}(\bm k) \cdot  \bm A(\bm r)] \right \} \,,
		\label{gnonlocal}
	\end{align}
	or, in an equivalent form more convenient for the calculations below, by
	\begin{align}
			g^{(1)}(\bm k, \bm r) = \frac{2 \pi e^2}{\hbar c^2} \delta({\varepsilon_{\bm k}^g} - \hbar \omega) 
			\sum_{\alpha = x,y} \nabla_\alpha \, {\rm Im}
			\bigl \{ [\bm v_{cv}(\bm k) \cdot  \bm A(\bm r)]^* \partial_{k_{\alpha}}
			[\bm v_{cv}(\bm k, \bm k') \cdot  \bm A(\bm r)] \bigr \}  \nonumber \\
			- \frac{2 \pi e^2}{\hbar c^2} \delta(\varepsilon_{\bm k}^g  - \hbar \omega) \sum_{\alpha = x,y} 
			\partial_{k'_\alpha}  {\rm Im}
			\bigl \{[\bm v_{cv}(\bm k, \bm k') \cdot  \bm A(\bm r)]^* \,  
			\nabla_\alpha \, [\bm v_{cv}(\bm k, \bm k') \cdot  \bm A(\bm r)] \bigr \}
			\,  \nonumber \\
			+ \frac{2 \pi e^2}{c^2}  \delta'(\varepsilon_{\bm k}^g  - \hbar \omega)\, {\rm Im}
			\left \{ [\bm v_{cv}(\bm k) \cdot  \bm A(\bm r)]^*  (\bm v_{\bm k}^v \cdot \nabla) \, [\bm v_{cv}(\bm k) \cdot  \bm A(\bm r)] \right \} \,,
			\label{gnonlocal_alt}
		\end{align}
\end{widetext}
where 
 $\mathrm{Im}\{..\}$ denotes the imaginary part of expression, $\delta'(\varepsilon) = d \delta (\varepsilon) / d \varepsilon$, 
$\bm v_{\bm k}^v = (1/\hbar) d \varepsilon_{\bm k}^v / d \bm k$, $\varepsilon_{\bm k}^g = \varepsilon_{\bm k}^c - \varepsilon_{\bm k}^v$, and one should set $\bm k' = \bm k$ after the differentiation.

The term $g^{(1)}(\bm k, \bm r)$ describes the non-locality of optical generation. It shows that spatially inhomogeneous radiation generates
asymmetric in the $\bm k$ space distribution of electrons. The asymmetry comes from the wave vector dependence of the 
matrix elements of the interband velocity [first two lines in Eq.~\eqref{gnonlocal_alt}] and the band dispersions 
[third line in Eq.~\eqref{gnonlocal_alt}].
For the exemplary case of $\bm A(\bm r) = \bm A \exp [i \phi(\bm r)]$, 
Eq.~\eqref{gnonlocal_alt} yields 
\begin{multline}
	g^{(1)}(\bm k, \bm r) = - \frac{2 \pi e^2}{\hbar c^2} \delta(\varepsilon_{\bm k}^g - \hbar \omega) \nabla \phi \cdot  \partial_{\bm k'} |\bm v_{cv} (\bm k, \bm k') \cdot \bm A|^2
	\,  \\
	+ \frac{2 \pi e^2}{c^2}  \delta'(\varepsilon_{\bm k}^g - \hbar \omega) (\nabla \phi \cdot \bm v_{\bm k}^v ) \,  |\bm v_{cv} (\bm k) \cdot \bm A|^2 \,
	 \,.
\end{multline}

\subsection{Summary}

Summarizing the calculations above, we conclude that the current density of particles in the conduction band induced by spatially inhomogeneous radiation to first order in $q/k$ has the form
\begin{equation}\label{j_general_final}
\bm j (\bm r) = \bm j^{(\rm loc)} (\bm r) + \bm j^{(\rm nl)} (\bm r)  \,, 
\end{equation}
where 
\begin{equation}\label{j_local_final} 
\bm j^{(\rm loc)} (\bm r) = \sum_{\bm k} \bm v_{\bm k}^{c} \, f(\bm k, \bm r) \,,
\end{equation}
$f(\bm k, \bm r)$ is the electron distribution function, which is found from the kinetic equation with the local generation term 
$g^{(0)}(\bm k, \bm r)$ given by Eq.~\eqref{glocal},
\begin{multline}\label{j_nonlocal_final}
\bm j^{(\rm nl)} (\bm r) =  \sum_{\bm k} \tau_1 g^{(1)}(\bm k, \bm r) \bm v_{\bm k}^{c}  \\
+ \frac{\i }{2}  \sum_{\bm k} \sum_{\alpha = x,y} \tau_1 \nabla_{\alpha} g^{(0)}(\bm k, \bm r) \,
 [(\partial_{k_\alpha} - \partial_{k'_\alpha}) \bm v_{\bm k, \bm k'}^{c} ]_{\bm k' = \bm k}  \,, 
\end{multline}
and $g^{(1)}(\bm k, \bm r)$ is the non-local correction to the generation term given by Eqs.~\eqref{gnonlocal} or~\eqref{gnonlocal_alt}.  
The current~\eqref{j_nonlocal_final} is given in the relaxation time approximation with the momentum relaxation time $\tau_1$.  
We note that the sum of two terms in Eq.~\eqref{j_nonlocal_final} is invariant with respect to the choice of the wave function phases
while the individual terms are not. The current $\bm j^{(\rm loc)}$ is determined by the spatial distribution of the intensity and polarization 
of radiation whereas the current $\bm j^{(\rm nl)}$ is also sensitive to the spatial structure of electromagnetic field phase.

Below we apply the developed theory to calculate the charge and spin-valley currents induced by structured light in 2D Dirac 
materials.

\section{2D Dirac material}\label{Sec_2D}

We consider the class of direct gap 2D materials described by the effective 
Hamiltonian~\cite{Castro-Neto2009,Kormanyos:2013,Bernevig:2006}
\begin{equation}\label{H0}
	\mathcal{H} = \hbar a(\nu\sigma_x k_x + \sigma_y k_y) + \delta\sigma_z \,,
\end{equation}
where $\sigma_{\alpha}\,(\alpha=x,y,z)$ are the Pauli matrices, $a$ and $\delta$ are the band-structure parameters, 
and $\nu = \pm 1$ is the spin (or valley) index. $\mathcal{H}$ with $\nu = \pm 1$ correspond to the blocks of the extended Hamiltonian
which describe the states with opposite spin projections or in different valleys related by time reversal symmetry.
The parameter $a$ determines the linear-in-$\bm k$ coupling between the valence and conduction band states
and $2 \delta$ is the band gap. 

The effective Hamiltonian~\eqref{H0} is relevant with varying accuracy to a wide class of 2D systems, 
including atomically thin crystals and quantum wells. In graphene, the states with $\nu = \pm 1$ belong to the opposite $K_+$ and $K_-$ valleys 
(which are additionally spin degenerate)~\cite{Castro-Neto2009}. In TMDC monolayers, the spin degeneracy in the valleys is lifted, so that the states 
with $\nu = \pm 1$ are also characterized by the opposite spin projections~\cite{Kormanyos:2013}. In narrow gap heterostructures, such as HgTe/CdHgTe quantum wells with close-to-critical thickness, the Dirac states are formed at the $\Gamma$ point of the 2D Brillouin zone and the index $\nu = \pm 1$ corresponds to the ''spin-up'' and ''spin-down'' blocks of the extended Hamiltonian~\cite{Bernevig:2006,Tarasenko:2015,Durnev:2022ac}.
	
The energy spectrum, velocity, and wave functions of the conduction and valence bands have the form
\begin{equation}
	\eps_{\bm k}^{c/v} = \pm \eps_{k} \,, \;\; \bm v^{c/v}_{\bm k}  = \pm \hbar a^2 \bm k / \eps_k \,,
\end{equation}
and 
\begin{equation}
		\label{psi_cv}
		\chi^{c/v}_{\nu \bm k} = \pm \frac{1}{\sqrt{2\eps_k(\eps_k \mp \delta)}} 
		\left[
		\begin{array}{c}
			- \hbar a (\nu k_x - i k_y) \\
			\delta \mp \eps_k
		\end{array}
		\right] \,, 
\end{equation}
respectively, where  $\varepsilon_k = \sqrt{\delta^2 + (\hbar a k)^2}$. 
	
The interband matrix elements of the velocity operator $\bm v_{cv}(\bm k, \bm k') = (\chi^c_{\nu \bm k})^\dag  \bm v \chi^v_{\nu \bm k'}$, 
which determine the generation terms~\eqref{glocal} and \eqref{gnonlocal}, and the intraband matrix elements
$\bm v^{c}_{\bm k, \bm k'} = (\chi^c_{\nu \bm k})^\dag \bm v \chi^c_{\nu \bm k'} $, which determine the current~\eqref{j_nonlocal_final},
are given by 
\begin{multline} \label{vcv}
		\bm v_{cv}(\bm k, \bm k') \cdot \bm A = \frac{\hbar a^2}{2 \sqrt{\eps_k \eps_{k'} (\eps_k - \delta)(\eps_{k'}+\delta)}} \\
		\times \bigl\{ \delta \bm A \cdot(\bm k + \bm k') + i \nu \left[\bm A \times (\bm k \eps_{k'} + \bm k' \eps_k) \right]_z  \\
		+ \bm A \cdot (\bm k \eps_{k'} - \bm k' \eps_k) + i \nu \delta \left[\bm A \times (\bm k - \bm k') \right]_z  \bigr\} \:,
\end{multline}	
and
\begin{multline} \label{vcc}
v_{\bm k, \bm k'}^{c}   = \frac{\hbar a^2}{2\sqrt{\eps_k \eps_{k'} (\eps_k - \delta)(\eps_{k'}-\delta)}} 
\bigl\{ \eps_k \bm k'+ \eps_{k'} \bm k \\ - (\bm k + \bm k') \delta
+ i \nu \left[ \eps_{k'} \bm k -  \eps_k \bm k' + (\bm k' - \bm k) \delta \right] \times \bm n \bigr\}\:,
\end{multline}
where $\bm n = (0,0,1)$ is the 2D normal.

Then, the local optical generation rates, Eq.~\eqref{glocal}, in the valleys assume the form
\begin{multline}\label{g_local_Dirac}
		g_{\nu}^{(0)} (\bm k, \bm r) = \frac{\pi e^2 a^2}{\hbar \omega^2 \varepsilon_{k}^2} \delta(2 \varepsilon_k - \hbar \omega) \bigl\{S_0 (\bm r)\left(\delta^2+\varepsilon_k^2\right)
		\\ - \hbar^2 a^2\bigl[\left(k_x^2-k_y^2\right) S_1 (\bm r) + 2 k_x k_y S_2 (\bm r)\bigr]\vphantom{{\varepsilon^c_{\bm k}}^2}
		+ 2 \nu \delta \eps_k S_3(\bm r) \bigr\}  \,,
\end{multline}
where 
$S_0 = |E_x|^2 + |E_y|^2$, $S_1 = |E_x|^2 - |E_y|^2$, $S_2 = E_xE_y^* + E_x^*E_y$, $S_3 = \i(E_xE_y^*-E_x^*E_y)$, and
$\bm E(\bm r) = (\i \omega/c)\bm A(\bm r)$ is the electric field amplitude. In the paraxial approximation for normally incident beam, 
i.e., when the in-plane photon wave vectors $q$ are much smaller than $\omega/c$, $S_j$ correspond to the (non-normalized) Stokes parameters. 
The parameter $S_0$ defines the local intensity of radiation, $S_1/S_0$ and $S_2/S_0$ are the degrees of linear polarization in the $(xy)$ frame and in the diagonal axes, respectively, and $S_3/S_0$ is the degree of circular polarization. 
The terms $\propto S_1, S_2$ in Eq.~\eqref{g_local_Dirac} describe the optical alignment of electron momenta, i.e., the creation of anisotropic distribution of electrons in the $\bm k$ space controlled by the linear polarization~\cite{Zakharchenya:1982,Merkulov:1991,Durnev2021b,Saroka:2022}. The term $\propto \nu S_3$ describes the optical orientation, i.e., the selectivity of optical transitions in the valleys (or in the spin subspaces) to the circular polarization~\cite{OO_book,Mak:2012}. 
Note that the band-edge absorption can be enhanced due to the Coulomb interaction between photoexcited electrons and holes~\cite{Leppenen:2021}.

\section{Electric and spin-valley currents}\label{Sec_currents}

Now we calculate the electric and spin-valley currents defined as 
\begin{eqnarray}
\bm j^e(\bm r) = e \kappa \sum_{\nu = \pm 1} \bm j_\nu  (\bm r) \,, \nonumber \\   
\bm j^{sv}(\bm r) = \kappa \sum_{\nu = \pm 1} \nu \bm j_\nu  (\bm r) \,,
\end{eqnarray}
where $\bm j_\nu$ are the flows of electrons in the valleys, $e$ is the electron charge, and $\kappa$ is the factor that takes into account a possible additional degeneracy of the Dirac states not included in the Hamiltonian~\eqref{H0}. In graphene, for instance, $\kappa = 2$ due to the spin degeneracy.

Before proceeding to the details of calculations, we make two comments.  First, we note that the mechanisms of the current generation are different for 
the contributions $\bm j^{(\rm loc)}$ and $\bm j^{(\rm nl)}$ given by the local and nonlocal generation terms, Eqs.~\eqref{j_local_final} and \eqref{j_nonlocal_final}, respectively. 

\begin{figure}
		\centering
		\includegraphics[width=\linewidth]{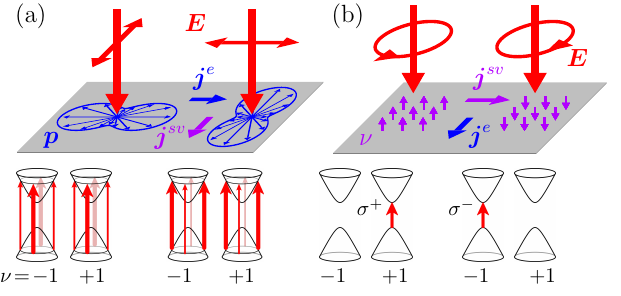}
		\caption{Generation of electric $\bm j^e$ and spin-valley $\bm j^{sv}$ currents by radiation with spatially varying polarization. 
		(a) Currents are driven by the alignment of electron momenta at interband optical transitions induced by linearly polarized light and 
		the subsequent momentum scattering. 
		(b) Currents are driven by the spin (or valley) polarization of electrons by circularly polarized light and the emerging spin (or valley) diffusion. 
		Electric and spin-valley currents are additionally coupled by the direct and inverse spin/valley Hall effects.}
		\label{fig_lin_circ}
\end{figure}

The currents $\bm j^{(\rm loc)}$ are formed in two steps. In the first step, the absorption of spatially inhomogeneous radiation leads to the emergence of photoelectrons with the nonuniform profile of the density, momentum alignment, or spin-valley polarization in the 2D plane. The subsequent relaxation of such 
a nonuniform distribution in the second step gives rise to direct currents determined by the gradients of these parameters.
For the radiation with nonuniform intensity, the major contribution to the electric current comes from the diffusion of photoinduced electrons towards the regions with lower electron density. This is the Dember photoelectric effect well known for 3D and 2D semiconductors~\cite{Dember1931}.  
The currents driven by the spatial gradients of linear and circular polarization originate mostly from the optical alignment and optical orientation, as illustrated in Fig.~\ref{fig_lin_circ}. The interband absorption of linearly polarized light results in anisotropic in the $\bm k$ space generation of electrons in the conduction band aligned perpendicular to the electric field polarization. Such a momentum alignment, shown by blue curves in Fig.~\ref{fig_lin_circ}(a), is the same in both valleys and stems from the dependence of the probability of optical transition on the relative orientation between the electron wave vector $\bm k$ and the electric field $\bm E$, see Eq.~\eqref{g_local_Dirac} and Refs.~\cite{Alperovich:1981a,Trushin:2011,Durnev2021b}. 
The scattering of electrons with spatially inhomogeneous momentum alignment reflecting the spatial variation of the linear polarization leads to an electric current $\bm j^e$. For the radiation with nonuniform circular polarization, Fig.~\ref{fig_lin_circ}(b), the spin-valley current $\bm j^{sv}$ comes from the local spin (or valley)  polarization of electrons created by circularly polarized light and the emerging spin (or valley) diffusion~\cite{Averkiev:1983a,Bakun:1984,Jin:2018}.

The current $\bm j^{(\rm nl)}$ given by the nonlocal contribution to the generation term originates from the asymmetric in the $\bm k$ space correction to the electron generation. It emerges simultaneously with the optical excitation and decays with the momentum relaxation time $\tau_1$. For the plane electromagnetic wave,
the asymmetry comes from the transfer of the in-plane photon momenta to the electron system, and $\bm j^{(\rm nl)}$ represents the photon drag current~\cite{Danishevskii1970, Gibson1970,Shalygin2007,Entin2010,Mikheev2018,Svintsov:2025}.
The currents $\bm j^{(\rm nl)}$ are typically small compared to the currents $\bm j^{(\rm loc)}$ by the factor $1/(k l)$, where $k$ is the electron wave vector and 
$l$ is the mean free path. However, they are sensitive to the phase structure of the electromagnetic field.

Second, the electric and spin-valley currents do not flow independently, they are converted to each other due to the spin/valley Hall effect (SHE/VHE) 
and the inverse spin/valley Hall effect (ISHE/IVHE)~\cite{dyakonov_book,Dyakonov:2007,Mak:2014}.
To first order in the charge-spin/valley conversion, the corrections to the currents have the form
\begin{eqnarray}
			\label{conversion} 
			\Delta \bm j^e &=& e \theta\, \bm n \times \bm j^{sv} \:, \nonumber \\ 
			\Delta \bm j^{sv} &=& (\theta/e) \, \bm n \times \bm j^e \:,
\end{eqnarray} 
where $\theta \ll 1$ is the spin/valley Hall angle. 
Theoretical calculations of the valley Hall angle in 2D materials have been attracting much attention~\cite{Ado:2015, Ando:2015, Kovalev:2018, Glazov:2020a}.
The valley Hall angle in the two-band Dirac model is estimated as $\theta \sim \hbar/(\delta \tau)$ for electrons at the conduction-band bottom in gapped Dirac materials, whereas $\theta$ vanishes in the gapless systems like graphene. In systems with spin-orbit interaction, a finite spin Hall angle can be expected even for 
gapless structures like HgTe/CdHgTe quantum wells of the critical thickness.

In the subsections below, we calculate the contributions $\bm j^{(\rm loc)}$ and $\bm j^{(\rm nl)}$ to electric and spin-valley currents.

\subsection{Electric currents caused by electron diffusion and optical alignment of electron momenta}\label{Subsec_j_e_loc}
	
The electric current given by the local generation term has the form, see Eq.~\eqref{j_local_final},
\begin{equation}
\bm j^e (\bm r) = e \kappa \sum_{\bm k} \bm v_{\bm k}^c f_e(\bm r, \bm k) \,,
\end{equation}
where $f_e = \sum_\nu f_\nu$. The distribution function $f_e$ is found from the kinetic equation
\begin{equation}\label{kinetic_f_e}
	\bm v^{c}_{\bm k}\cdot\nabla f_e(\bm k,\bm r) =  g_e^{(0)}(\bm k,\bm r) + I \{f_e(\bm k,\bm r)\} 
\end{equation}
with the local generation term $g_e^{(0)} = \sum_\nu g_\nu^{(0)}$. Equation~\eqref{g_local_Dirac} for $g_\nu^{(0)}$ yields 
\begin{multline}\label{g_e_local_explicit}
		g_e^{(0)} (\bm k, \bm r) = \frac{2\pi e^2 a^2}{\hbar \omega^2 \varepsilon_{k}^2} \delta(2 \varepsilon_k - \hbar \omega) \bigl\{S_0 (\bm r)\left(\delta^2+\varepsilon_k^2\right)
		\\ - \hbar^2 a^2\bigl[\left(k_x^2-k_y^2\right) S_1 (\bm r) + 2 k_x k_y S_2 (\bm r)\bigr]\vphantom{{\varepsilon^c_{\bm k}}^2}\bigr\} \,.
\end{multline}
The generation rate~\eqref{g_e_local_explicit} includes the polarization-independent contribution $\propto S_0$ and the terms $\propto (k_x^2-k_y^2)S_1$ and $\propto 2 k_x k_y S_2$ describing the local alignment of electron momenta by linearly polarized light, see Fig.~\ref{fig_lin_circ}(a). The alignment is the same in both Dirac cones whereas the contributions sensitive to circular polarization cancel out.
	
Multiplying Eq.~\eqref{kinetic_f_e} by $\bm v^c_{\bm k}$, summing up over the electron wave vector $\bm k$ and taking into account that $\sum_{\bm k} \bm v^c_{\bm k} g_e^{(0)} = 0$, we obtain
\begin{equation}\label{j_align1}
	\bm j^e(\bm r) = -e \kappa \sum_{\bm k} \tau_1 \bm v_{\bm k}^c \left( \bm v^{c}_{\bm k}\cdot\nabla f_e \right)\:,
\end{equation}
where $\tau_1$ is the momentum relaxation time defined by $\tau_1^{-1} = -\langle \bm v I \{ f_e \} \rangle / \langle \bm v f_e \rangle$
and the angular brackets denote averaging over directions of the wave vector $\bm k$.

Equation~\eqref{j_align1} for the current already contains the spatial derivative. Therefore, the distribution function $f_e$ can be calculated 
in the local response approximation, i.e., from the equation $g_e^{(0)} + I \{f_e\} = 0$. The generation term $g_e^{(0)}$ contains the zeroth 
and the second angular harmonics in the $\bm k$ space and so does the distribution function $f_e$.
The current sensitive to the linear polarization is determined by the second angular harmonic which decays with the relaxation time
$\tau_2$ defined by $\tau_2^{-1} = -\langle v_x v_y I\{f_e\}\rangle/\langle v_x v_y f_e \rangle$. 
It gives $\langle (v_{\alpha} v_{\beta} - \delta_{\alpha \beta}v^2/2) f_e \rangle = \tau_2 \langle (v_{\alpha} v_{\beta} - \delta_{\alpha \beta}v^2/2) g_e^{(0)} \rangle$. The zeroth angular harmonic of the distribution function relaxes with the energy relaxation time $\tau_\epsilon$ (photoelectrons relax to the band bottom) and then, at longer times, decays with the photoelectron lifetime $\tau_0$ determined mainly by electron-hole recombination. To describe the diffusion of photoelectrons we introduce the diffusion length $L_e$ by $\sum_{\bm k} \tau_1 (v^2/2) f_e = L_e^2 \sum_{\bm k} g_e^{(0)}$. Then, calculating the sum in Eq.~\eqref{j_align1}, we finally obtain
\begin{multline}\label{j_align2}
	\bm 	j^e = - \frac{\kappa e^3 L_e^2}{8 \hbar \, \delta \, w^3} (w^2 + 1)  \, \nabla S_0 \\
	+ \frac{\kappa e^3 \tau_1 \tau_2 a^2}{32 \hbar \,\delta \,w^5 }(w^2-1)^{2} \left( \tilde{\nabla} S_1 + \bm n \times \tilde{\nabla} S_2 \right) \:,
\end{multline}
where $w = \hbar \omega/(2 \delta) \geq 1$ is the dimensionless frequency and $\tilde{\nabla} = (\nabla_x, -\nabla_y, 0)$.

Equation~\eqref{j_align2} is relevant to the first order in the gradients of the Stokes parameters. The corresponding criteria for the first and second terms 
in Eq.~\eqref{j_align2} are that the scale of field variation $L$ is larger than the diffusion length $L_e$ and the mean free path $l$, respectively.

\subsection{Electric current due to nonlocal generation}\label{Subsec_j_e_nonloc}

The currents determined by the nonlocal contribution to the generation, Eq.~\eqref{j_nonlocal_final}, are formed directly at optical transitions and decay with the momentum relaxation time. Here, additional scattering processes are not required for the current formation. Taking into account the equations for the generation terms,
Eqs.~\eqref{glocal} and~\eqref{gnonlocal_alt}, the matrix elements of the velocity operator derived in Sec.~\ref{Sec_2D}, and summing up the contributions 
to the current from two valleys we obtain after some algebra 
\begin{multline} \label{j_nonlocal2}
		\bm j^e = \frac{\kappa e^3\tau_1 a^2}{64\,\delta^2\, w^6}  \biggl\{ - 2w^2(w^2+3)\,\bm n\times\nabla S_3 \\
		+2\left[(w^2+1)^2 + (w^4-1)(\ln \tau_1)' \right]\, {\rm Im}\bigl(E^*_x \nabla E_x + E^*_y \nabla E_y\bigr) \\+ (w^2-1)^2\left[1 - (\ln \tau_1)' \right]\,{\rm Im}\bigl[\bm E^* (\nabla \cdot \bm E) -\bm E^*\times(\nabla \times \bm E)\bigr]\biggr\} .
\end{multline}
Here, we take into account a possible dependence of the momentum relaxation time on electron energy, which results in additional corrections 
$\propto (\ln \tau_1)' = \eps (d \tau_1/d\eps)/\tau_1$. For $\tau_1 \propto \eps^\alpha$, one has $(\ln \tau_1)' = \alpha$.

The electric current~\eqref{j_nonlocal2} contains the contribution driven by the gradient of the circular polarization ($\propto \nabla S_3$) and the contributions that are sensitive to  the phase and polarization of electromagnetic field
and cannot be expressed in terms of spatial derivatives of the Stokes parameters. For the field in the form 
$\bm E(\bm r) = \bm E \exp[\i \varphi (\bm r)]$ with the fixed vector $\bm E$, the phase-sensitive contributions assume the form
\begin{multline}\label{j_nonlocal2a} 
	\bm j^e = \frac{\kappa e^3\tau_1 a^2}{64\,\delta^2\, w^6}  \biggl\{2\left[(w^2+1)^2 + (w^4-1)(\ln \tau_1)' \right]\,S_0\nabla\varphi \\+ (w^2-1)^2\left[1 - (\ln \tau_1)' \right]\,\left( S_1\tilde{\nabla}\varphi + S_2\bigl(\bm n \times \tilde{\nabla}\varphi\bigr) \right)\biggr\} \,.
\end{multline}
The current~\eqref{j_nonlocal2a} corresponds to the photon drag and contains the longitudinal contribution along $\nabla \varphi$ and the transverse contribution perpendicular to $\nabla \varphi$. The latter is controlled by the linear polarization and vanishes for unpolarized light.

\subsection{Spin-valley currents caused by diffusion of optically polarized electrons}\label{Subsec_j_vs_loc}
	
The spin-valley currents given by the local generation term emerge in the case of radiation with spatially varying Stokes parameter $S_3(\bm r)$,
e.g., radiation with the nonuniform profile of circular polarization. Microscopically, the currents result from the selective excitation of electrons in 
the $\nu = 1$ and $\nu = -1$ states by circularly polarized light and the subsequent diffusion of spin-valley polarized electrons, see Fig.~\ref{fig_lin_circ}(b).

The spin-valley current is given by
\begin{equation}
\bm j^{sv} (\bm r) = \kappa \sum_{\bm k} \bm v_{\bm k}^c f_{sv}(\bm r, \bm k) \,,
\end{equation}
where $f_{sv} = \sum_\nu \nu f_\nu$ is the distribution function satisfying the kinetic equation
\begin{equation}\label{kinetic_f_sv}
	\bm v^{c}_{\bm k}\cdot\nabla f_{vs}(\bm k,\bm r) =  g_{sv}^{(0)}(\bm k,\bm r) + I \{f_{vs}(\bm k,\bm r)\}  \,,
\end{equation}
with the local generation term $g_{sv}^{(0)} = \sum_\nu \nu g_\nu^{(0)}$,
\begin{equation}\label{gs_local_explicit}
		g_{sv}^{(0)} (\bm k, \bm r) = \frac{4 \pi e^2 a^2\delta}{\hbar \omega^2 \varepsilon_{k}}\,\delta(2 \varepsilon_k - \hbar \omega) S_3 (\bm r) \,.
\end{equation}
Note that, in the Dirac model, the spin-valley polarization by circular polarized light occurs only in materials with a finite band gap $2\delta$,
and $g_{sv}^{(0)}$ vanishes in gapless materials with $\delta = 0$.

After the calculations similar to those described in the Sect.~\ref{Subsec_j_e_loc} we obtain
\begin{equation}\label{jv_diffusion3}
		\bm j^{sv} = -\frac{\kappa e^2 L_{sv}^2}{4 \hbar\,\delta\,w^2} \,  \nabla S_3 \:,
\end{equation}
where $L_{sv}$ is the spin-valley diffusion length defined by $\sum_{\bm k} \tau_1 (v^2/2) f_{sv} = L_{sv}^2 \sum_{\bm k} g_{sv}^{(0)}$.
Note that the spin-valley diffusion length is limited not only by the lifetime of photoelectrons $\tau_0$ but also the spin-valley relaxation time $\tau_{sv}$,
$L_{sv}^2 \propto \tau_0 \tau_{sv}/(\tau_0 + \tau_{sv})$.

\subsection{Spin-valley currents due to nonlocal generation}\label{Subsec_j_vs_nonloc}

The spin-valley current given by the nonlocal generation term is calculated similarly to the corresponding electric current, see Sec.~\ref{Subsec_j_e_nonloc}. 
The calculation gives
\begin{multline}\label{js_nonlocal2}
		\bm j^{sv} = -\frac{\kappa e^2\tau_1 a^2}{32\,\delta^2w^5}  \biggl\{(3w^2+1)\, \bm n \times \nabla S_0  \\
		-(w^2-1)\,\left( \bm n \times \tilde{\nabla} S_1 - \tilde{\nabla} S_2\right)  \\ 
		+2\left[(w^2+1) + (w^2-1) (\ln \tau_1)' \right] \\
		\bm n\times {\rm Re}\bigl[\bm E^* (\nabla \cdot \bm E) - (\bm E^* \cdot \nabla)  \bm E \bigr] \biggr\}\:.
\end{multline}

The spin-valley current~\eqref{js_nonlocal2} contains the contributions driven by the gradient of the radiation intensity $S_0$ and the gradients of the Stokes parameters $S_1$ and $S_2$ determining the linear polarization. Besides, it contains the contribution sensitive to the electromagnetic field phase. For the field 
$\bm E(\bm r) = \bm E \exp[\i \varphi(\bm r)]$ with fixed $\bm E$, this contribution assumes the form
\begin{equation}
		\bm j^{sv} = \frac{\kappa e^2\tau_1 a^2}{16\,\delta^2w^5} \left[(w^2+1) + (w^2-1) (\ln \tau_1)' \right] S_3 \nabla \varphi  \,.
\end{equation} 
It can be viewed as the spin-valley circular photon drag.

Concluding Sec.~\ref{Sec_currents}  we make an additional note. Electric signals measured in experiments are contributed by the currents of electrons in the conduction band 
and holes in the valence band. The contributions of electrons and holes do not generally compensate each other due to electron-hole asymmetry
in the dispersion or in relaxation times. In doped structures, Coulomb interaction between photoexcited and resident carriers can play an important role in the formation of a net electric current. Particularly, if the photoexcited electrons and holes flow in the same direction, their charge currents would compensate each other. However, in fact, momentum transfer from the photoexcited electron and holes to the resident carriers results in the formation of the direct flow of resident carriers and, hence, the net charge current.

\section{Symmetry analysis of the currents}\label{Sec_symmetry}

\begingroup
\renewcommand{\arraystretch}{2.3}
\begin{table*}[!ht]
	\begin{center}
		\begin{tabular}{c c c c c c}
			\hline
			\hline
			\,\\[-3em] & Electric currents & \begin{picture}(0,0)\put(-2,2){$\theta$}\put(-8.5,-3){$\longleftrightarrow$}\end{picture} & Spin-valley currents & & Realization\\
			\hline  
			$Q_1$ \vline & $- \dfrac{\kappa e^3 L_e^2}{8 \hbar \, \delta \, w^3} (w^2 + 1)$  & \vline $~W_1$ \vline & $-\dfrac{\kappa e^2\tau_1 a^2}{32\,\delta^2w^5} (3w^2+1)$ & \vline & $\propto \nabla S_0$ \\
			$Q_2$ \vline & $\dfrac{\kappa e^3\tau_1 \tau_2 a^2}{32 \hbar \,\delta \,w^5 }(w^2-1)^{2}$ & \vline $~W_2$ \vline & $\dfrac{\kappa e^2\tau_1 a^2}{32\,\delta^2w^5} (w^2-1)$ & \vline & $\propto \nabla S_{1,2}$ \\
			$Q_3$ \vline & $\dfrac{\kappa e^3\tau_1 a^2}{32\,\delta^2\, w^4} (w^2+3)$  & \vline $~W_3$ \vline & $-\dfrac{\kappa e^2 L_{sv}^2}{4 \hbar\,\delta\,w^2}$ & \vline & $ \propto\nabla S_3$ \\
			$Q_4$ \vline & $\dfrac{\kappa e^3\tau_1 a^2}{32\,\delta^2\, w^6} \left[(w^2+1)^2 + (w^4-1)(\ln \tau_1)' \right]$ & \vline $~W_4$ \vline & $0$ & \vline & $ \propto S_0\nabla\varphi$ \\
			$Q_5$ \vline & $\dfrac{\kappa e^3\tau_1 a^2}{64\,\delta^2\, w^6} (w^2-1)^2\left[1 - (\ln \tau_1)' \right]$ & \vline $~W_5$ \vline & $0$ & \vline & $\propto S_{1,2}\nabla\varphi$ \\
			$Q_6$ \vline & $0$ & \vline $~W_6$ \vline & $ -\dfrac{\kappa e^2\tau_1 a^2}{16\,\delta^2w^5} \left[(w^2+1) + (w^2-1) (\ln \tau_1)' \right] $ & \vline & $\propto S_3\nabla\varphi$ \\[1ex]
			\hline 
			\hline
		\end{tabular}
	\end{center} 
	\caption{Summary of microscopic calculations of electric and spin-valley currents. The table presents analytical results for the parameters $Q_j$ and $W_j$ introduced by Eqs.~\eqref{Q123} and~\eqref{Q456} and Eqs.~\eqref{W123} and~\eqref{W456}, respectively, $w = \hbar \omega/2 \delta$ is the photon energy normalized by the band gap $2\delta$. Electric and spin-valley currents are additionally interconverted via the spin/valley and the inverse spin/valley Hall effects, see Eq.~\eqref{conversion}, which is quantified by the spin/valley Hall angle $\theta$.
		}
	\label{tab1}
\end{table*}
\endgroup	

The electric and spin-valley currents calculated above can be systematized on the symmetry basis.
We assume that the electromagnetic field varies smoothly in the 2D plane and consider the photocurrents linear in the field gradients. 
The most general expressions for such electric and spin-valleys currents are
	\begin{eqnarray}
		j^e_\alpha = \sum_{\beta \gamma \delta} Q_{\alpha \beta \gamma \delta} E_{\beta}^* \nabla_\gamma E_\delta + \mathrm{c.c.}\:, \\
		j^{sv}_\alpha = \sum_{\beta \gamma \delta} W_{\alpha \beta \gamma \delta } E_{\beta}^* \nabla_\gamma E_\delta + \mathrm{c.c.}\:,
	\end{eqnarray}
where $Q$ and $W$ are the fourth-rank (complex) tensors, the indices $\alpha,\,\beta,\, \gamma$, and $\delta$ run over the in-plane axes, 
and $\bm E(\bm r)$ is the complex electric field amplitude. 

In isotropic 2D systems, there are three (complex) combinations $\bm E^* (\nabla \cdot \bm E)$, $(\bm E^* \cdot \nabla)  \bm E$, and $E^*_x \nabla E_x + E^*_y \nabla E_y$ that transform as polar vectors. The real and imaginary parts of these terms (or their superpositions) determine 6 linearly independent contributions to the electric current. Hence, the general expression for the electric current is 
\begin{equation}
\bm j^e = \bm j_{\rm pol}^e + \bm j_{\rm ph}^e \,,
\end{equation}
where
	\begin{multline}
		\label{Q123}
		\bm j_{\rm pol}^e = 2 Q_1\, {\rm Re}\bigl(E^*_x \nabla E_x + E^*_y \nabla E_y\bigr)  \\
		+ 2 Q_2\, {\rm Re}\bigl[\bm E^* (\nabla \cdot \bm E) -\bm E^*\times(\nabla \times \bm E)\bigr] \\
		+ 2 Q_3\, {\rm Im}\bigl[\bm E^* (\nabla \cdot \bm E) - (\bm E^* \cdot \nabla)  \bm E \bigr] \:,
	\end{multline}
	\begin{multline}
		\label{Q456}
		\bm j_{\rm ph}^e = Q_4 \, {\rm Im}\bigl(E^*_x \nabla E_x + E^*_y \nabla E_y\bigr)  \\ 
		+ Q_5\, {\rm Im}\bigl[\bm E^* (\nabla \cdot \bm E) -\bm E^*\times(\nabla \times \bm E)\bigr] \\
		+ Q_6\,{\rm Re}\bigl[\bm E^* (\nabla \cdot \bm E) - (\bm E^* \cdot \nabla)  \bm E \bigr] \:,
	\end{multline}
and $Q_j$ ($j = 1..6$) are real parameters. 
In deriving the equations above, we used the equality $\bm E^*\times(\nabla \times \bm E) = E^*_x \nabla E_x + E^*_y \nabla E_y - (\bm E^* \cdot \nabla)  \bm E$.
	
The current $\bm j_{\rm pol}^e$ can be equivalently written via the spatial gradients of the Stokes parameters as follows
	\begin{multline}
		\label{Q123_Stokes}
		\bm j_{\rm pol}^e = Q_1 \nabla S_0 + Q_2(\tilde{\nabla} S_1 + \bm n \times \tilde{\nabla} S_2) - Q_3\, \bm n \times \nabla S_3 \;.
	\end{multline}
We remind that $S_0 = |E_x|^2 + |E_y|^2$, $S_1 = |E_x|^2 - |E_y|^2$, $S_2 = E_xE_y^* + E_x^*E_y$, $S_3 = \i(E_xE_y^*-E_x^*E_y)$,  $\tilde{\nabla} = (\nabla_x, -\nabla_y, 0)$, and $\bm n = (0,0,1)$ is a unit vector normal to the 2D plane. So, the current $\bm j_{\rm pol}^e$ is completely determined by the profiles of the light intensity 
and the Stokes polarization parameters.  

The current $\bm j_{\rm ph}^e$ is determined not only by the polarization but also by the phase of the field. 
To explore it, we consider the exemplary field in the form $\bm E(\bm r) = \bm E \exp [i \phi(\bm r)]$, i.e., the field with the fixed amplitude 
$\bm E$ and spatially varying phase $\phi(\bm r)$. In this special case, $\bm j_{\rm ph}^e$ is proportional to the phase gradient $ \nabla \phi$ and reads
	\begin{equation} 
		\label{j_PDE}
		\bm j^e_{\rm ph} = Q_4 S_0 \nabla \phi +  Q_5 \left(S_1 \tilde{\nabla} \phi +  S_2 \bm n \times \tilde{\nabla} \phi \right) + Q_6 S_3 \bm n \times \nabla \phi\:.
	\end{equation} 
For a single plane wave with the in-plane wave vector $\bm q$, $\nabla \phi = \bm q$, and the current $\bm j_{\rm ph}^e$ is proportional to $\bm q$. 
Therefore, Eq.~\eqref{j_PDE} can be viewed as a generalization of the photon drag current to the electromagnetic field with arbitrary varying phase.  	
We stress that Eq.~\eqref{Q456} is more general and is reduced to Eq.~\eqref{j_PDE} only for special cases. 

Analogously, there are three (complex) combinations $(\bm n \times \bm E^*) (\nabla \cdot \bm E)$, $(\bm E^* \cdot \nabla) (\bm n \times  \bm E)$, and $\bm n \times (E^*_x  \nabla E_x + E^*_y \nabla E_y)$ that transform as axial vectors. Therefore, the spin-valley current is also described phenomenologically by 6 real parameters $W_j$ as follows
\begin{equation}
\bm j^{sv} = \bm j^{sv}_{\rm pol} + \bm j^{sv}_{\rm ph} \,,
\end{equation}
where 
\begin{equation}
		\label{W123}
		\bm j^{sv}_{\rm pol} = W_1\,\bm n \times \nabla S_0 + W_2 \left( \bm n \times \tilde{\nabla} S_1 - \tilde{\nabla} S_2\right)
		+ W_3 \nabla S_3 \,,
\end{equation}
\begin{multline}
		\label{W456}
		\bm j^{sv}_{\rm ph} = W_4 \, \bm n \times {\rm Im}\bigl(E^*_x \nabla E_x + E^*_y \nabla E_y\bigr)  \\ 
		+ W_5\, \bm n \times {\rm Im}\bigl[\bm E^* (\nabla \cdot \bm E) -\bm E^*\times(\nabla \times \bm E)\bigr] \\
		+ W_6\, \bm n \times {\rm Re}\bigl[\bm E^* (\nabla \cdot \bm E) - (\bm E^* \cdot \nabla)  \bm E \bigr] \:.
\end{multline}

Time reversal symmetry imposes additional restrictions on the $Q_j$ and $W_j$ parameters. $Q_1$, $Q_2$, and $Q_6$ as well as $W_3$, $W_4$, and $W_5$ do not change sign under time reversal, which suggests that they are proportional to dissipation parameters, such as $\tau$, in even powers. In contrast, $Q_3$, $Q_4$, and $Q_5$ as well as $W_1$, $W_2$, and $W_6$ change their signs under time reversal and, therefore, are proportional to dissipation parameters in odd powers.
	
In Sec.~\ref{Sec_currents} we have calculated different contributions to the electric and spin-valley currents. In Sec.~\ref{Sec_symmetry} we have developed their phenomenological description. The results of both sections are summarized in Tab.~\ref{tab1}, which shows the calculated expressions for the parameters $Q_j$ and $W_j$.
The results are presented for 2D Dirac materials with the band gap $2\delta$. For gapless materials, the results are obtained considering the asymptotic 
$\delta \rightarrow 0$. For the opposite case of optical excitation at the band edge, 
$w= \hbar\omega / (2\delta)$ is close to unity and one can express the currents via the effective mass $m^* = \delta/a^2$.

\section{Results and discussion}\label{Sec_Results}

\begin{figure*}
\includegraphics[width=0.92\linewidth]{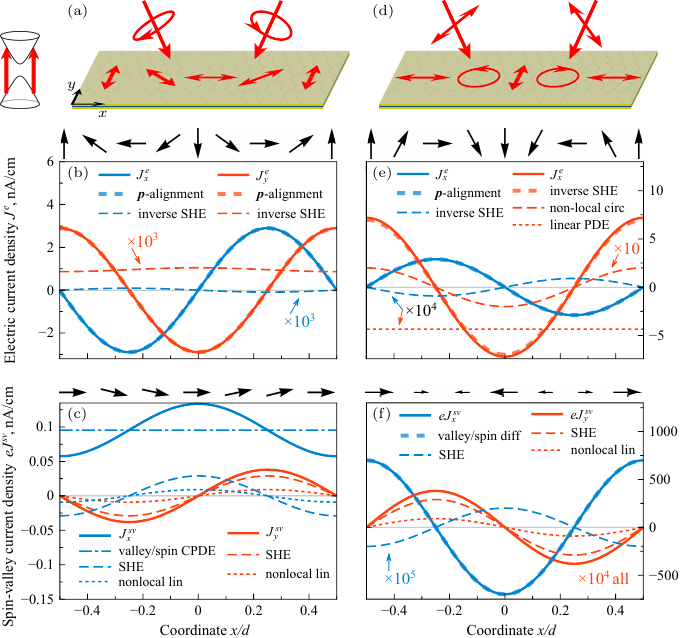}
\caption{Spatial distributions of electric and spin-valley currents in a gapped Dirac material subject to 1D optical gratings.
(a,d) Polarization patterns in the optical gratings formed by (a) two circularly polarized waves with opposite helicity and (d) 
two linearly polarized waves with orthogonal polarization. 
Panels (b,c) and (e,f) present the current distributions for the optical gratings (a) and (d), respectively.
Solid curves show the total current projections, whereas dashed, dotted, and dash-dotted curves depict partial contributions of different mechanisms. 
Black arrows over each panel schematically show the directions of the corresponding currents. 
The distributions are plotted for the radiation intensity $I = 1\text{ W/cm}^2$, the photon energy $\hbar \omega = 2.2$\,eV, the band gap $2\delta=2$\,eV, 
the in-plane wave vector $q = 0.1(\omega/c)$ corresponding to the grating period $d  \approx 2.8\text{ }\mu\text{m}$, the velocity band parameter $a=c/300$,
the relaxation times $\tau_1=\tau_2=1\text{ ps}$, the spin/valley diffusion length $L_{sv}=1\,\mu\text{m}$, and the spin/valley Hall angle $\theta=10^{-2}$.}
		\label{tmdc_plots}
\end{figure*}

Now, we illustrate the developed theory on the example of
optical fields in the form of 1D polarization gratings, Fig.~\ref{tmdc_plots}.
Such gratings are formed by two obliquely incident plane waves with orthogonal polarizations with the resulting electric field in the 2D plane 
\begin{equation}
\label{grating} 
	\bm E(x) = \frac{E_0}{\sqrt{2}} \left( \bm e_1\e^{\i q x} + \bm e_2 \e^{-\i q x} \right) \,,
\end{equation}	
where $E_0$ is the field amplitude (uniform over the grating), $q$ is the in-plane wave vector, $\bm e_1$ and $\bm e_2$ are the complex polarization vectors. 
In the paraxial approximation, one can disregard the difference between the polarization vectors and their 2D projections onto the sample plane.

We consider two types of polarization grating. The first one is formed by circularly polarized waves with the polarization vectors $\bm e_{1,2}=(1,\mp \i)/\sqrt{2}$, Fig.~\ref{tmdc_plots}(a). The resulting electric field $\bm E(x) =  E_0 (\cos qx,\,\sin qx)$ is linearly polarized and its polarization varies along $x$ with the spatial period $d = \pi/q$. The corresponding Stokes parameters are $S_0 = E_0^2$, $S_1(x) = E_0^2 \cos (2 qx)$, $S_2(x) = E_0^2 \sin(2qx)$, and $S_3 = 0$. 
The electric and spin-valley currents are found from Eqs.~\eqref{Q123} and~\eqref{Q456} and Eqs.~\eqref{W123} and~\eqref{W456}, respectively, with 
additional contributions due to the spin/valley and the inverse spin/valley Hall effects given by Eq.~\eqref{conversion}. This gives the electric currents
\begin{eqnarray}
\label{Je_grating1}
	J^{e}_x &=& 2 q E_0^2 \left[- Q_2 \sin(2qx) + e \theta W_2 \sin(2qx) \right] , \\
	J^{e}_y &=& 2q E_0^2\left\{ Q_2 \cos(2qx) - e \theta \left[W_2  \cos(2qx) - W_6/2 \right] \right\} \nonumber ,
\end{eqnarray}
and the spin-valley currents
\begin{eqnarray}
\label{Jv_grating1}
	J^{sv}_{x} &=& -2q E_0^2 \left[W_2 \cos(2qx) - W_6/2 + (\theta/e) Q_2  \cos(2qx) \right] , \nonumber \\
	J^{sv}_{y} &=& -2 q E_0^2 \left[W_2 \sin(2qx) + (\theta/e) Q_2 \sin(2qx) \right]  .
\end{eqnarray}

The second type of grating is formed by two linearly polarized waves with the polarization vectors $\bm e_{1,2}=(\pm 1,1)/\sqrt{2}$, so that the resulting electric field takes the form $\bm E(x) = E_0 (\i \sin qx,\,\cos qx)$, see Fig.~\ref{tmdc_plots}(d). The polarization pattern of this radiation is the alternation of $\sigma_+$ and $\sigma_-$ circular polarizations with intermediate linear polarizations along $x$ or $y$ axes. The corresponding Stokes parameters are $S_0 = E_0^2$, 
$S_1(x) = -E_0^2 \cos (2 qx)$, $S_2 = 0$, and $S_3(x) = -E_0^2 \sin(2 q x)$. Such an electromagnetic field drives the electric currents
\begin{eqnarray}
\label{Je_grating2}
	J^{e}_x &=& 2 q E_0^2 \left[Q_2 \sin(2qx) - e \theta W_2 \sin(2qx) \right] ,  \\
	J^{e}_y &=& 2 q E_0^2 \left[ Q_3 \cos(2qx) + Q_5/2 - e \theta W_3 \cos(2qx) \right] \nonumber ,
\end{eqnarray}
and the spin-valley currents
\begin{eqnarray}
\label{Jv_grating2}
	J^{sv}_{x} &=& -2 q E_0^2 \left\{W_3 \cos(2qx) + (\theta/e) \left[ Q_3 \cos(2qx) + Q_5/2 \right] \right\} , \nonumber \\
	J^{sv}_{y} &=& 2 q E_0^2 \left[W_2 \sin(2qx) + (\theta/e) Q_2 \sin(2qx) \right] .
\end{eqnarray}

Figure~\ref{tmdc_plots} shows the spatial distributions of electric and spin-valley currents induced by the optical gratings discussed above in a gapped 2D Dirac material.
Solid curves show the total current densities while dashed, dotted, and dash-dotted curves present the contributions of different mechanisms. 
The distributions are plotted 
for the parameters relevant to TMDC monolayers. 
Due to strong spin-orbit splitting, the states in the $K_+$ and $K_-$ valleys ($\nu = \pm 1$, respectively) are characterized by the opposite spin projections. 
Therefore, valley currents in TMDC monolayers are also spin currents.

For the linearly polarized optical grating,
Fig.~\ref{tmdc_plots}(b), the electric current is dominated by the contribution driven by the polarization gradients:
$J_x^e \propto \nabla_x S_1$, $J_y^e \propto \nabla_x S_2$. This contribution oscillates in space with the grating period $d$.
Microscopically, it  comes mostly from the optical alignment of electron momenta (the $Q_2$ parameter). 
In TMDCs, the optical alignment is not very efficient since it occurs only due to the band non-parabolicity, and the $Q_2$  parameter contains the
small factor $(w^2 - 1)^2 \approx 4 (\hbar\omega - 2\delta)^2/(2\delta)^2$ at $\hbar\omega - 2\delta \ll 2 \delta$. 
The spin-valley current, Fig.~\ref{tmdc_plots}(c), also contains the space-oscillating contribution driven by the polarization gradients. It originates from the the non-locality of optical generation 
(the $W_2$ term) and the conversion from the electric current (the $\theta Q_2$ term). The magnitude of the spin-valley current is small compared to that of the electric current since both $W_2/Q_2$ and $\theta$ are proportional to the small parameter $(\omega \tau)^{-1} \sim 10^{-3}$.

\begin{figure*}
\includegraphics[width=0.85\linewidth]{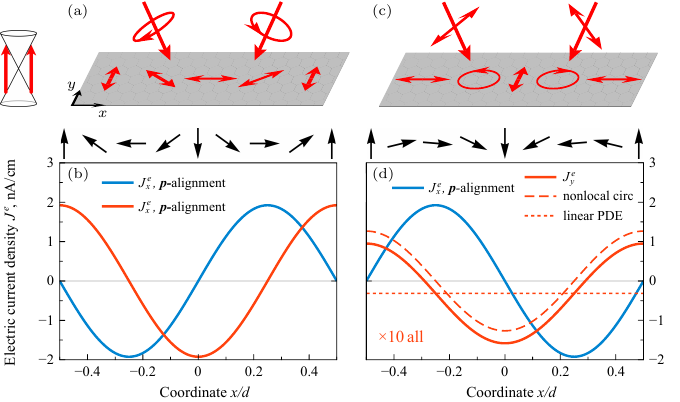}
\caption{Spatial distributions of electric current density in graphene subject to 1D optical gratings formed by two circularly polarized waves with opposite helicity (a, b) or
by two linearly polarized waves with orthogonal polarization (c, d). Solid curves show the projections of the total electric current, whereas dashed and dotted curves depict partial contributions of different mechanisms. Black arrows over each panel schematically show the current directions.  
The distributions are plotted for the radiation intensity $I = 1\text{ W/cm}^2$, the photon energy $\hbar \omega = 0.2$\,eV, 
the in-plane wave vector $q = 0.1(\omega/c)$ corresponding to the grating period $d  \approx 31$~$\mu$m, the electron velocity $a=c/300$ and
the relaxation times $\tau_1=\tau_2=0.1\text{ ps}$ relevant to graphene.}
\label{graphene_plots}
\end{figure*}

Interestingly, the spin-valley current contains a significant contribution, which is uniform throughout the grating and flows along the $x$ axis, see the blue dash-dotted line in Fig.~\ref{tmdc_plots}(c). This contribution is described by the parameter $W_6$, see Eq.~\eqref{W456}, and essentially represents the spin-valley circular photon drag.
To shed light on its nature, we remind that the considered optical grating is formed by two plane waves with the opposite helicity and the opposite in-plane wave vectors, see Eq.~\eqref{grating}. Due to the sensitivity of optical transitions in the valleys to the particular circular polarizations and the transfer of in-plane photon wave vectors to electrons, these two waves induce oppositely directed photon drags in the valleys. This results in the spatially uniform contribution to the spin-valley current in Fig.~\ref{tmdc_plots}(c). Note that the uniform contribution to the electric current coming from the inverse spin/valley Hall effect (the $\theta W_6$ parameter) is also present in Fig.~\ref{tmdc_plots}(b). However, it is small compared to the spatially oscillating current.

The distributions of electric and spin-valley currents for the optical grating of second type are plotted in Figs.~\ref{tmdc_plots}(e, f). 
Here, the overwhelming contribution to the spin-valley current is $J_x^{sv} \propto W_3 \nabla_x S_3$ which originates from the diffusion of spin-valley polarized 
electrons from the regions where the local field is circularly polarized.  This current is determined by the polarization pattern and oscillates with the spatial period $d$, see Fig.~\ref{tmdc_plots}(f). Concerning the electric current, Fig.~\ref{tmdc_plots}(e), the $J_y^e$ component is manly determined by the conversion of the large spin-valley current via the inverse spin/valley Hall effect whereas $J_x^e$ stems from the optical alignment of electron momenta in the regions where the field contains linearly polarized component. Other contributions to the electric and spin-valley currents include those related to the non-locality of optical generation.

The smaller, but still important, contribution to the electric current is the phase and polarization sensitive contribution $\propto Q_5$, which leads for this optical grating
to a uniform current flowing along $y$, see the red dotted line in Fig.~\ref{tmdc_plots}(e). This current originates from the transverse photon drag in the direction perpendicular to the light incident plane and controlled by the linear polarization, see Eq.~\eqref{j_PDE}. Indeed, the grating is formed by two waves with the orthogonal linear polarizations and the opposite in-plane wave vectors. The photon drag current $J_y^e \propto q _x S_2$, however, is the same for both individual waves,  giving rise to a net uniform electric current.

Figure~\ref{graphene_plots} shows the spatial distributions of electric currents induced by the optical gratings discussed above in a gapless 2D Dirac material with the parameters relevant to graphene. At zero band gap $2\delta$, the photoresponses in both valleys coincide in the isotropic approximation and all the parameters $W_j$ in Tab.~\ref{tab1} vanish. The valley Hall effect in gapless graphene is also absent~\cite{Ado:2015}. Fig.~\ref{graphene_plots} reveals that the 
dominant contribution to the electric current originates from the optical alignment of electron momenta, which is particularly strong in Dirac materials 
with the linear energy dispersion. Other contributions, which occur for the second optical grating, are related to the non-locality of optical generation and the linear photon drag. The role of these contributions is enhanced in graphene in the infrared spectral range as compared to TMDCs in the optical spectral range, since 
the parameter $\omega \tau$ gets smaller.

\section{Summary}\label{Sec_Summary}

To summarize, we have studied the generation of electric and spin-valley currents in 2D systems at interband absorption of light with structured intensity, polarization, and phase. Using the density matrix formalism and the Wigner transformation, we have derived the quasi-classical rate of optical transitions in the momentum and coordinate spaces, which goes beyond the Fermi golden rule. The optical generation contains the local and non-local terms. The former is determined by the local intensity and polarization of the electromagnetic field whereas the latter is sensitive to the spatial gradients of the field, including the polarization and phase gradients. 

We have revealed the mechanisms of the charge and spin-valley current generation, which are different for the contributions $\bm j^{(\rm loc)}$ and $\bm j^{(\rm nl)}$ given by the local and nonlocal generation terms, respectively. The currents $\bm j^{(\rm loc)}$ are typically stronger. They occur due to the relaxation of nonuniform  distribution of the electron density, momentum alignment, or spin-valley polarization in the 2D plane created by structured radiation.
The currents $\bm j^{(\rm nl)}$ originate from the asymmetry of optical transitions in the $\bm k$ space and emerge simultaneously with the optical excitation. 
Based on symmetry classification, we have established that electric and spin-valley currents induced by structured light in isotropic 2D systems
are described in the paraxial approximation by two sets of six linearly independent parameters. Using the kinetic theory, we have obtained analytical expressions 
for those contributions in 2D Dirac materials.

The developed theory has been applied to the study of spatial texture of electric and spin-valley currents emerging in TMDC layers and graphene under excitation with polarization gratings formed by two coherent optical fields with orthogonal polarizations. In this geometry, the currents controlled by the gradients of the Stokes polarization parameters oscillate in space with the grating period. 
Additionally, the electric and spin-valley currents contain contributions, which are uniform throughout the grating. These contributions are sensitive to both the polarizations and the phases of the incident waves and represent the charge and spin-valley photon drag. Our results suggest that radiation with structured parameters can be used for the efficient injection and control of electric and spin-valley currents in 2D crystals.

\acknowledgments 
	This work was supported by the Russian Science Foundation (Project No. 22-12-00211-$\Pi$). A.A.G. and M.V.D. also acknowledge the support from the Basis Foundation for the Advancement of Theoretical Physics and Mathematics.

\bibliographystyle{apsrev4-1-customized}
\bibliography{bibliography}

\end{document}